\documentclass[aps,twocolumn,showpacs]{revtex4}
\begin{document}
\draft
\title{Bose-Einstein Condensation in the Framework of
\mbox{\boldmath$\kappa$}-Statistics}
\author{A. Aliano\footnote{Electronic address: aliano@polito.it},
G. Kaniadakis\footnote{Electronic address: kaniadakis@polito.it
(corresp. author, fax: ++39 011 5647399)} and E.
Miraldi\footnote{Electronic address: miraldi@polito.it}}
\address{Dipartimento di Fisica and INFM, Politecnico di Torino,
Corso Duca degli Abruzzi 24, 10129 Torino, Italy}
\begin {abstract}
In the present work we study the main physical properties of a
gas of $\kappa$-deformed bosons described through the statistical
distribution function $f_{\scriptstyle
\kappa}=Z^{-1}[\exp_{{\scriptstyle
\{\kappa\}}}(\beta(\frac{1}{2}m\mbox{\boldmath
$v$}^2-\mu))-1]^{-1}$. The deformed $\kappa$-exponential
$\exp_{{\scriptstyle \{\kappa\}}}(x)$, recently proposed in Ref.
[G.Kaniadakis, Physica A {\bf 296}, 405, (2001)], reduces to the
standard exponential as the deformation parameter
$\kappa\rightarrow0$, so that $f_0$ reproduces the Bose-Einstein
distribution. The condensation temperature $T_c^{^{\,
{\scriptstyle\kappa}}}$ of this gas decreases with increasing
$\kappa$ value, and approaches the $^{4\!}H\!e(I)$ -
$^{4\!}H\!e(II)$ transition temperature $T_{\lambda}=2.17K$,
improving the result obtained in the standard case ($\kappa=0$).
The heat capacity $C_V^{^{\, {\scriptstyle\kappa}}}(T)$ is a
continuous function and behaves as $B_{\scriptstyle \kappa}
T^{3/2}$ for $T< T_c^{^{\,\scriptstyle \kappa}}$, while for $T >
T_c^{^{\, {\scriptstyle\kappa}}}$, in contrast with the standard
case $\kappa=0$, it is always increasing.
\end {abstract}
 \maketitle
\section{Introduction}
The liquid $^{4\!}H\!e$ behaviour at low temperature and pressure
has been studied in the approximation of the standard ideal boson
gas, and described by the Bose-Einstein (BE) statistical
distribution:
\begin{equation}
f=\frac{1}{\exp(\epsilon)-1} \ \ ,\label{1}
\end{equation}
where $\epsilon=\beta\left(\frac{1}{2}m\mbox{\boldmath
$v$}^2-\mu\right)$. The condensation temperature, calculated
within this model, has the value $T_{c}=3.07 K$, while the
fluid-superfluid transition temperature for the $^{4\!}H\!e$,
measured at the saturation pressure, is $T_{\lambda}=2.17 K$.
Therefore the difference between the two results is about forty
percent. Moreover the heat capacity behavior presents in the two
cases some qualitative and quantitative differences. Some models
depending on a free parameter based on the group theory,
describing the liquid $^{4\!}H\!e$ behavior, have been proposed
in Ref.s \cite{g1,g2,g3,g4}. Recently the liquid $^{4\!}H\!e$ is
considered in the framework of the nonextensive statistical
mechanics \cite{3,4,5}. In Ref. \cite{6}, a model is developed for
the liquid $^{4\!}H\!e$, starting from the nonextensive quantum
statistical distribution of Ref.s \cite{7,8,9}, given by:
\begin{equation}
f=\frac{1}{[\exp_q(-\epsilon)]^{-1}-1} \ \ ,\label{2}
\end{equation}
where $\exp_q(x)=[1+(1-q)x]^{\frac{1}{1-q}}$. The condensation
temperature $T_c^{^q}$ obtained in Ref. \cite{6} by using the Eq.
(\ref{2}), decreases with increasing $q$ value, becoming quite
similar to the experimental result.

In the present effort we propose a study using a new theoretical
model, in which the liquid $^{4\!}H\!e$ is viewed as a particles
gas described by a deformed Bose-Einstein statistical
distribution, recently proposed in Ref. \cite{10,11}, depending on
one continuous parameter $\kappa$, given by:
\begin{equation}
f=\frac{1}{\exp_{{\scriptstyle \{\kappa\}}}(\epsilon)-1} \ \ ,
\label{3}
\end{equation}
being
\begin{equation} \exp_{{\scriptstyle
\{\kappa\}}}(x)=\left(\sqrt{1+\kappa^2x^2}+\kappa
x\right)^{1/\kappa} \ \ .\label{4}
\end{equation}

The paper is organized as follows: in the Sect. II, after
recalling the main mathematical properties of
$\exp_{{\scriptstyle \{\kappa\}}}(x)$, we define the entropy which
gives the distribution given by Eq. (\ref{3}) using the maximum
entropy principle. In Sect. III we introduce a family of
integrals for defining the physical properties of the gas and
subsequently we calculate the particle number and energy. In
Sect. IV the condensation temperature and heat capacity of the
system are analyzed, while in the Sect. V some conclusions are
reported.

\section{The \mbox{\boldmath
$\kappa$}-statistics} In Ref. \cite{10} it has been shown that the
deformed exponential $\exp_{_{\{{\scriptstyle \kappa}\}}}(x)$
given by Eq. (\ref{4}) satisfies the equation:
\begin{equation}
\exp_{_{\{{\scriptstyle \kappa}\}}}(x) \exp_{_{\{{\scriptstyle
\kappa}\}}}(-x)=1 \ \ . \label{5}
\end{equation}
It is easy to verify that the most general solution of the Eq.
(\ref{5}) is given by $\exp_{_{\{{\scriptstyle \kappa,g}
\}}}(x)\,\,=\,\, \exp_{_{\{{\scriptstyle \kappa}\}}}(g(\kappa
x)/\kappa)$, where the generator $g(x)$ of the
$\exp_{_{\{{\scriptstyle \kappa,g} \}}}(x)$ is an arbitrary, odd
and increasing function, behaving as
$g(x){\atop\stackrel{\textstyle\sim}{\scriptstyle x\rightarrow
0}}x$. The choice $g(x)=x$ reduces to the $\exp_{_{\{{\scriptstyle
\kappa}\}}}(x)$ while if we set
$g(x)=\sinh(x)$ we obtain the standard exponential $\exp(x)$.\\
We remark that $\exp_{_{\{{\scriptstyle \kappa}\}}}(x)$ reduces
to $\exp(x)$ as $\kappa$ approaches to zero. Namely
$\exp_{_{\{{\scriptstyle 0}\}}}(x)=\exp (x)$ where
$\exp_{_{\{{\scriptstyle 0}\}}}(x)\equiv \lim_{\kappa\rightarrow
0}\,\exp_{_{\{{\scriptstyle \kappa}\}}}(x)$. It is interesting to
note that $\exp_{_{\{{\scriptstyle \kappa}\}}}(x)$ satisfies
several properties of $\exp(x)$, some of these properly deformed.
For instance $\exp_{_{\{{\scriptstyle \kappa}\}}}(x)$ is a
positive monotonic increasing function $\forall x\in {\bf R}$,
symmetric with respect to the parameter $\kappa$, being
$\exp_{_{\{{\scriptstyle -\kappa}\}}}(x)=\exp_{_{\{{\scriptstyle
\kappa}\}}}(x)$. Asymptotically the $\exp_{_{\{{\scriptstyle
\kappa}\}}}(x)$ follows the power law given by
\begin{eqnarray}
\exp_{_{\{{\scriptstyle
\kappa}\}}}(x){\atop\stackrel{\textstyle\sim}{\scriptstyle
x\rightarrow \pm\infty}}(2|\kappa x|)^{\pm1/|\kappa|} \ \ .
\label{6}
\end{eqnarray}
Others useful relations obtained in Ref. \cite{10} are: for
$\forall a \in {\bf R}$ it is verified that
$\exp_{_{\{{\scriptstyle \kappa}\}}}(ax)=[\exp_{_{\{{\scriptstyle
a\kappa}\}}}(x)]^{\,a}$ and holds the relation:
\begin{equation}
\exp_{_{\{{\scriptstyle \kappa}\}}}(x) \exp_{_{\{{\scriptstyle
\kappa}\}}}(y) =\exp_{_{\{{\scriptstyle \kappa}\}}}( x
\oplus\!\!\!\!\!^{^{^{\, {\scriptstyle\kappa}}}}\,\,y )\ \ ,
\end{equation}
where the $\kappa$-sum $x \oplus\!\!\!\!\!^{^{\scriptstyle
\kappa}}\,\,y$, is defined through $x
\oplus\!\!\!\!\!^{^{\scriptstyle
\kappa}}\,\,y=x\sqrt{1+\kappa^2y^2}+y\sqrt{1+\kappa^2x^2}$
\cite{10,11,12}.

Moreover it is remarkable that the inverse function of the
$\kappa$-exponential, namely $\kappa$-logarithm, is defined by
\begin{equation}
\ln_{_{\{{\scriptstyle \kappa}\}}}(x)=
\frac{x^{\kappa}-x^{-\kappa}}{2\kappa} \ \ . \label{8}
\end{equation}
It is a monotonic increasing function $\forall x\in{\bf R}^+$ and
reduces to the standard logarithm when $\kappa\rightarrow0$. The
asymptotic behaviour of the $\ln_{_{\{{\scriptstyle
\kappa}\}}}(x)$ follows again a power law given by
\begin{equation}
\ln_{_{\{{\scriptstyle
\kappa}\}}}(x){\atop\stackrel{\textstyle\sim}{\scriptstyle
x\rightarrow +\infty}}\frac{1}{2|\kappa|}\, x^{|\kappa|}\ \ \ \ \
; \ \ \ \ \ \ln_{_{\{{\scriptstyle
\kappa}\}}}(x){\atop\stackrel{\textstyle\sim}{\scriptstyle
x\rightarrow {\,0^+}}}-\frac{1}{2|\kappa|}\,
\frac{1}{x^{|\kappa|}} \ \ .
\end{equation}
Others properties of the $\ln_{_{\{{\scriptstyle \kappa}\}}}(x)$
can be found in Ref. \cite{10}. It is well known that the maximal
entropy principle asserts that for a given entropy
\begin{equation}
S_{\scriptstyle \kappa}= \int_{\cal R}\ d^nv \,\,
\sigma_{\scriptstyle \kappa} (f) \ \ , \label{10}
\end{equation}
the most probable distribution for the system (formed by brownian
particles interacting with a bath) is obtained as solution of the
following variational equation:
\begin{equation}
\delta\left[ S_{_{\scriptstyle \kappa}} -\beta\int_{\cal R}d^nv
\frac{1}{2}m\mbox{\boldmath$v$}^2\, f +\beta\mu\int_{\cal R} d^n
v\, f \right] =0 \ \ ,\label{11}
\end{equation}
where the constants $\beta$ and $\beta\mu$ are the Lagrange
multipliers.
\\ In Ref. \cite{10} it has been postulated for the local entropy $\sigma_{\scriptstyle
\kappa} (f)$ the following expression:
\begin{equation} \sigma_{\scriptstyle \kappa}
(f)=-\int df \ln_{_{\{{\scriptstyle \kappa}\}}}(f)  \ \
,\label{12}
\end{equation}
where $\ln_{_{\{{\scriptstyle \kappa}\}}}(f)$ is defined through
 Eq. (\ref{8}). Then, from Eq. (\ref{11}) the classical
distribution follows
\begin{equation}
f_{\scriptstyle \kappa}=\exp_{_{\{{\scriptstyle
\kappa}\}}}\left(-\beta\left(\frac{1}{2}m\mbox{\boldmath$v$}^2-\mu\right)\right)
\ \ . \label{13}
\end{equation}
We note that for $\kappa\rightarrow0$ the entropy
$S_{\scriptstyle \kappa}$ reduces to the well known
Shannon-Boltzmann-Gibbs entropy and the distribution of Eq.
(\ref{13}) becomes the MB one.

In this work we are interested in obtaining $\kappa$-modified BE
statistical distribution. Therefore, we can consider the local
entropy having the form
\begin{equation}
\sigma_{\scriptstyle \kappa} (f)=-\int df \ln_{_{\{{\scriptstyle
\kappa}\}}}\left(\frac{f}{f+1}\right) \ \ , \label{14}
\end{equation}
that reproduces for $\kappa=0$ the standard boson local entropy.
By solving Eq. (\ref{11}) the stationary distribution derived from
Eq. (\ref{14}) assumes the form:
\begin{equation}
f_{\scriptstyle \kappa}=\frac{1}{\exp_{{\scriptstyle
\{\kappa\}}}\left(\beta\left(\frac{1}{2}m\mbox{\boldmath$v$}^2\!-\!\mu\right)\right)-1}
\ \ . \label{15}
\end{equation}
The Eq. (\ref{15}) is the BE $\kappa$-modified that we are looking
for and reduces to the BE stationary distribution as $\kappa$
approaches to 0.
\section{Ideal gas of \mbox{\boldmath
$\kappa$}-deformed bosons} Let us study an ideal gas formed by
$N$ identical particles governed by the statistical distribution
of Eq. (\ref{15}). For calculating the typical thermodynamic
properties of this gas, we will use in the following the
integrals $J^{^{\, {\scriptstyle\kappa}}}_n(x)$ defined by
\begin{equation}
J^{^{\, {\scriptstyle\kappa}}}_n(x)={\int_0^\infty
\frac{t^n}{\exp_{_{\{{\scriptstyle \kappa}\}}}(t+x)-1}dt} \ \ ,
\label{16}
\end{equation}
which converge when $-1<n<1/|\kappa|-1$.\\
It is easy to verify, taking into account Eq. (\ref{16}), that
\begin{equation}
\frac{dJ^{^{\, {\scriptstyle\kappa}}}_n (x)}{dx} = n\, J^{^{\,
{\scriptstyle\kappa}}}_{n-1}(x) \ \ . \label{17}
\end{equation}
The asymptotic behaviour of the integrals $J^{^{\,
{\scriptstyle\kappa}}}_n(x)$ for $x\rightarrow \infty$ is
obtained immediately using Eq. (\ref{6}):
\begin{equation}
J_n^{^{\,{\scriptstyle
\kappa}}}(x){\atop\stackrel{\textstyle\sim}{\scriptstyle
x\rightarrow  \infty}}
\frac{\Gamma(n+1)}{(2|\kappa|)^{1/|\kappa|}}\frac{\Gamma(1/|\kappa|-n-1)}{\Gamma(1/|\kappa|)}\,\,
x^{n+1-1/|\kappa|} \ \ . \label{18}
\end{equation}
Since it will be useful later on, we introduce the following
auxiliary function:
\begin{equation}
F_n^{^{\,{\scriptstyle \kappa}}}(x)=\frac{J_n^{^{\,
{\scriptstyle\kappa}}}(x)}{J_{n-1}^{^{\,
{\scriptstyle\kappa}}}(x)} \ \ , \label{19}
\end{equation}
for which the asymptotic behaviour for $\kappa\neq0$ assumes the
form
\begin{equation}
F_n^{^{\,{\scriptstyle \kappa}}}(x){\atop\stackrel{\textstyle\sim}
{\scriptstyle x\rightarrow  \infty}}\,\,
\frac{n|\kappa|}{1-(n+1)|\kappa|}x \ \ . \label{20}
\end{equation}
On the other hand, in the case $\kappa=0$, we have
\begin{equation}
F_n^{^{\,{0}}}(x){\atop\stackrel{\textstyle\sim}{\scriptstyle
x\rightarrow \infty}}\,\,  n \ \ . \label{21}
\end{equation}

Let us now consider the physical quantities in the thermodynamic
limit that one can immediately evaluate from the knowledge of the
distribution function defined through Eq.(\ref{15}); the total
particle number $N=N_{0}^{^{\, {\scriptstyle \kappa}}} +
\int_{_{-\infty}}^{^{+\infty}} f(v)d^3v$ is given by:
\begin{equation}
N = N_{0}^{^{\, {\scriptstyle \kappa}}}
+\frac{V}{2\pi^2}\left(\frac{2m}{\hbar^2}\right)^\frac{3}{2}\beta^{-\frac{3}{2}}
 J^{^{\, {\scriptstyle\kappa}}}_{\scriptscriptstyle 1/2}(\nu) \ \ ,\label{22}\\
\end{equation}
where $\nu=\beta|\mu|$, while $N_{0}^{^{\, {\scriptstyle\kappa}}}$
is the particle number on the ground state $\epsilon=0$:
\begin{equation}
\,\,\,\,\,\,\,\,\,\,\,\,\,\,\,\,\,\,\,\,\, N_{0}^{^{\,
\scriptstyle \kappa}}=\frac{1}{\exp_{_{\{{\scriptstyle
\kappa}\}}}(\nu)-1} \ \ .\label{23}\\
\end{equation}
For the total kinetic energy $U^{^{\, {\scriptstyle\kappa}}}=
\int_{_{-\infty}}^{^{+\infty}} \frac{1}{2}m\mbox{\boldmath
$v$}^2\, f(v)d^3v$ we obtain
\begin{equation}
U^{^{\, {\scriptstyle\kappa}}}
=\frac{V}{2\pi^2}\left(\frac{2m}{\hbar^2}\right)^\frac{3}{2}\beta^{-\frac{5}{2}}
J^{^{\, {\scriptstyle\kappa}}}_{\scriptscriptstyle 3/2}(\nu) \ \ .
\label{24}
\end{equation}
Finally from Eq.s (\ref{22}) and (\ref{24}) we find
straightforwardly:
\begin{equation}
U^{^{\, {\scriptstyle\kappa}}} = (N - N_{0}^{^{\, {\scriptstyle
\kappa}}}) k_{B}T F_{\scriptscriptstyle 3/2}^{^{\,
{\scriptstyle\kappa}}}(\nu) \ \ .\label{25}
\end{equation}

\section{Physical properties of ideal \mbox{\boldmath
$\kappa$}-deformed boson gas} The condensation temperature
$T_{c}^{^{\, {\scriptstyle\kappa}}}$ for the $\kappa$-deformed
boson gas can be calculated by observing that at this temperature
the potential $\nu$ is equal to zero and all the $N$ particle are
in the excited states. Then from Eq.(\ref{22}) we have:
\begin{equation}
\frac{V}{2\pi^2}\left(\frac{2m}{\hbar^2}\right)^\frac{3}{2}(\kappa_{B}T_{c}^{^{\,
{ \scriptstyle \kappa}}})^{\frac{3}{2}}J^{^{\, {\scriptstyle
\kappa}}}_{\scriptscriptstyle 1/2}(0)= N \ \ .\label{26}
\end{equation}
In the limit $\kappa\rightarrow0$, we obtain the standard
condensation temperature $T_c^{^{\scriptstyle 0}}$. Thus, by
taking into account that $J^{^{\, 0}}_{\scriptscriptstyle
1/2}(0)=2.612\sqrt \pi/2$ and choosing $m=m_{H\!e^4}$, Eq.
(\ref{26}) yields the value $T_c^{^{\, {0}}}=3.07K$. After these
considerations it is easy to find from Eq. (\ref{26}) the relation
existing between the $\kappa$-modified $T_c^{^{\,
{\scriptstyle\kappa}}}$ and the standard $T_c^{^{\, {0}}}$ one:
\begin{equation}
\frac{T_{c}^{^{\, {\scriptstyle\kappa}}}}{T_{c}^{^{\, {0}}}} =
\left[\frac{J^{^{\, {0}}}_{\scriptscriptstyle 1/2}(0)}{J^{^{\,
{\scriptstyle \kappa}}}_{\scriptscriptstyle
1/2}(0)}\right]^{\frac{2}{3}}.\\\label{27}
\end{equation}
From Eq. (\ref{26}) we note that $T_c^{^{\,
{\scriptstyle\kappa}}}$ decreases with increasing $\kappa$ value;
so by increasing the $\kappa$ value we could approach the
condensation temperature $T_c^{^{\, {\scriptstyle\kappa}}}$ to
the $^{4\!}H\!e$ experimental transition temperature, namely
$T_{\lambda}=2.17 K$.

Now we consider the behaviour of the particle ratio $N_{0}^{^{\,
{\scriptstyle\kappa}}}/N$ in the ground state, as a function of
the temperature for different $\kappa$ values. This number is
different from zero only for $T<T_{c}^{\scriptstyle \kappa}$ and
after recalling Eq.s (\ref{22}) and (\ref{26}) we obtain:
\begin{equation}
\frac{N_{0}^{^{\, {\scriptstyle\kappa}}}}{N} =
\left[1-\left(\frac{T}{T_{c}^{^{\, {\scriptstyle\kappa}}}}
\right)^\frac{3}{2} \right] \ \ .
\end{equation}
This expression of $N_{0}^{^{\, {\scriptstyle\kappa}}}/N$ can also
be written as
\begin{equation}
\frac{N_{0}^{^{\, {\scriptstyle\kappa}}}}{N} =
\left[1-\frac{J^{^{\, {\scriptstyle\kappa}}}_{\scriptscriptstyle
1/2}(0)}{J^{^{\, {0}}}_{\scriptscriptstyle
1/2}(0)}\left(\frac{T}{T_{c}^{^{\, {0}}}} \right)^\frac{3}{2}
\right] \ \ . \label{29}
\end{equation}

In Fig.1  the particle rate is plotted at the ground state
$N_0^{^{\, {\scriptstyle \kappa}}}/N$ vs the temperature rate
$T/T_c^{^{\, {0}}}$ for different $\kappa$ values. The shift of
the curves $N_0^{^{\, {\scriptstyle \kappa}}}/N$ towards left is
due to the decreasing of $T_c^{^{\, {\scriptstyle\kappa}}}$ as
$\kappa$ increases. This behaviour of $T_c^{^{\,
{\scriptstyle\kappa}}}$ is plotted in the insert of the fig.1.
Besides it is remarkable that this decrease of $T_c^{^{\,
{\scriptstyle\kappa}}}$ allows us to approach the experimental
liquid $^{4\!}H\!e$ transition temperature $T_{\lambda}=2.17K$.
In fact for $\kappa=0.39$ we obtain $T^{^{\,
{\scriptstyle\kappa}}}_c=2.70K$ (excess of 25\%) while for
$\kappa=0$ (standard bosons) we have $T_c^{^{\, {0}}}=3.07K$
(excess of 41\%). Finally we calculate the heat capacity
$C_{V}^{^{\, {\scriptstyle\kappa}}}$ the $\kappa$-deformed bosons
by using its definition
\begin{equation}
C_{V}^{^{\, {\scriptstyle\kappa}}} \doteq
\left(\frac{\partial U^{^{\, {\scriptstyle\kappa}}}}{\partial
T}\right)_{V}\ \ .\\\label{30}
\end{equation}
For obtaining the expression of $C_{V}^{^{\,
{\scriptstyle\kappa}}}$ we distinguish the two following cases:
\begin{eqnarray}
(I)\,\,\,\, T\leq T^{^{\,{\scriptstyle
\kappa}}}_c\,\,\,\,\,\Rightarrow\,\,\,\,\,\, (\nu=0)
\,\,\,\,\,\,,\,\,\,\,\,\,  N_{0}^{^{\, {\scriptstyle\kappa}}}\neq
0
\\\label{31} (II)\,\,\,\, T>T^{^{\, {\scriptstyle
\kappa}}}_c\,\,\,\,\,\Rightarrow\,\,\,\,\, (\nu\neq0)
\,\,\,\,\,\,,\,\,\,\,\,\,  N_{0}^{^{\, {\scriptstyle\kappa}}}=0
\label{32}
\end{eqnarray}
For case (I), after substituting Eq. (\ref{24}) into Eq.
(\ref{30}) we have:
\begin{equation}
\frac{C_{V}^{^{\, {\scriptstyle\kappa}}}}{k_{B} N} =
\frac{5}{2}F_{\scriptscriptstyle 3/2}^{^{\, {\scriptstyle
\kappa}}}(0)\left(\frac{T}{T_{c}^{^{\, {\scriptstyle
\kappa}}}}\right)^{\frac{3}{2}}.\\\label{33}
\end{equation}
\\ On the other hand, for case (II), preliminary we differentiate Eq. (\ref{24}) obtaining:
\begin{equation}
\frac{d U^{^{\, {\scriptstyle\kappa}}}}{dT} = k_{B} N\left[F^{^{\,
{\scriptstyle\kappa}}}_{\scriptscriptstyle 3/2}(\nu) +
T\frac{d\nu}{dT}\frac{dF_{\scriptscriptstyle 3/2}^{^{\, {
\scriptstyle \kappa}}}(\nu)}{d\nu}\right].\label{34}
\end{equation}
We note that $F^{^{\, {\scriptstyle \kappa}}}_{\scriptscriptstyle
3/2}(\nu)$ is given by Eq. (\ref{19}), while $dF^{^{\,
{\scriptstyle \kappa}}}_{\scriptscriptstyle 3/2}(\nu)/d\nu$ can
be calculated by combining Eq.s (\ref{19}) and (\ref{17}). Finally
the quantity $Td\nu/dT$ can be obtained by observing that the
total particle number $N$ given by Eq. (\ref{22}) is a constant
and then we have $dN/dT=0$. This last relation yields:
\begin{equation}
T\frac{d\nu}{dT}=-3 F^{^{\,
{\scriptstyle\kappa}}}_{\frac{1}{2}}(\nu) \ \ , \label{35}
\end{equation}
thus we can write $C_{V}^{^{\, {\scriptstyle\kappa}}}$ in its
definitive form in the case $T>T_{c}^{^{\,
{\scriptstyle\kappa}}}$:
\begin{equation}
\frac{C_{V}^{^{\, {\scriptstyle\kappa}}}}{k_{B} N} = \frac{5}{2}
F^{^{\, {\scriptstyle\kappa}}}_{\scriptscriptstyle
3/2}(\nu)-\frac{9}{2} F^{^{\,
{\scriptstyle\kappa}}}_{\scriptscriptstyle 1/2}(\nu) \ \
.\label{37}
\end{equation}
Finally  we can write the general expression of the specific heat
as follows:
\begin{eqnarray}
\frac{C_{V}^{^{\, {\scriptstyle\kappa}}}}{k_{B} N} = \left\{
\begin{array}{c} {\displaystyle\frac{5}{2}F^{^{\, {\scriptstyle
\kappa}}}_{\scriptscriptstyle 1/2}(0)\left(\frac{T}{T_{c}^{^{\,
{\scriptstyle \kappa}}}}\right)^{\frac{3}{2}}}
\,\,\,\,\,\,\,\,\,\,\,\,\,\, ; \hspace{.7in} \mbox{ $T\leq
T_{c}^{^{\, {\scriptstyle\kappa}}}$};
\\ \\ {\displaystyle \frac{5}{2}
F^{^{\, {\scriptstyle\kappa}}}_{\scriptscriptstyle
3/2}(\nu)-\frac{9}{2} F^{^{\,
{\scriptstyle\kappa}}}_{\scriptscriptstyle 1/2}(\nu)}\,\,\, ;
\hspace{.7in} \mbox{ $T>T_{c}^{^{\, {\scriptstyle\kappa}}}$}.
\label{38}
\end{array}\right.
\end{eqnarray}

It is clear that for studying $C_V^{^{\, {\scriptstyle\kappa}}}$
as a function of $T$ one needs to know $\nu=\nu(T)$ for
$T>T_{c}^{^{\, {\scriptstyle\kappa}}}$. In order to calculate
this function we can combine Eq.s (\ref{22}) and (\ref{26}):
\begin{equation}
\frac{T}{T_{c}^{^{\scriptstyle
\kappa}}}=\left[\frac{J_{\scriptscriptstyle 1/2}^{^{\,
{\scriptstyle \kappa}}}(0)}{J_{\scriptscriptstyle 1/2}^{^{\,
{\scriptstyle \kappa}}}(\nu)}\right]^\frac{2}{3} \ \ . \label{36}
\end{equation}
Eq. (\ref{36}) defines implicitly $\nu(T)$ and numerically one can
see that $\nu(T)$ is an increasing function. This function can be
obtained also directly from Eq. (\ref{35}) by performing an
integration.

In Fig.2 the specific heat $C_{V}^{^{\,
{\scriptstyle\kappa}}}/Nk_{B}$ vs the temperature $T/T_c^{^0}$
for different values of $\kappa$ is plotted. All curves are
continuous, while for $T=T_{c}^{^{\, {\scriptstyle\kappa}}}$ we
obtain a discontinuity in the first derivative which is due to
the phase transition. $C_{V}^{^{\, {\scriptstyle\kappa}}}$, for
$T<T_{c}^{^{\, {\scriptstyle\kappa}}}$, assumes the form
$C_{V}^{^{\, {\scriptstyle\kappa}}}=B_{\scriptstyle
\kappa}T^{3/2}$ where $B_{\scriptstyle \kappa}$ is a constant
depending on $\kappa$ and its behaviour is the standard power law
of the undeformed bosons. On the other hand, for $T>T_{c}^{^{\,
{\scriptstyle\kappa}}}$ we have that $C_{V}^{^{\,
{\scriptstyle\kappa}}}$ depends strongly on $\kappa$ and
asymptotically it tends to the values $3/2Nk_B$ only when
$\kappa=0$. For $\kappa\neq0$ we have that $C_{V}^{^{\,
{\scriptstyle\kappa}}}\rightarrow\infty$ for $T\rightarrow\infty$.
Moreover we can observe that the angular point, corresponding to
the transition from the non condensate to condensate state, is
shifted to the left as $\kappa$ increases, in accordance with the
behaviour of $T_c^{^{\, {\scriptstyle\kappa}}}$ vs $\kappa$.
Finally we consider the classical limit $T\rightarrow\infty$.
This limit is equivalent to the limit $\nu\rightarrow\infty$ and
for $\kappa=0$ we obtain the well known result:
\begin{equation}
C_{V}^{^{\, {0}}} {\atop\stackrel{\textstyle\sim}{\scriptstyle
\nu\rightarrow \infty}} \frac{3}{2}\,\, N k_{B} \ \ .
\end{equation}
On the other hand, for $\kappa\neq0$, after substituting into Eq.
(\ref{38}) the asymptotic behaviour of $F^{^{\,
{\scriptstyle\kappa}}}_{\scriptscriptstyle n}(\nu)$ given by Eq.
(\ref{20}), we have
\begin{equation}
C_V^{^{\, {\scriptstyle
\kappa}}}{\atop\stackrel{\textstyle\sim}{\scriptstyle
\nu\rightarrow \infty}}  \,\, \frac{3}{2}\,\, N k_{B}\,\,
\frac{4|\kappa|}{(2-5|\kappa|)(2-3|\kappa|)}\,\, \nu(T) \ \ .
\end{equation}

In conclusion we remark that the usual thermodynamic formalism of
standard bosons, properly modified, can be adopted entirely to
treat also the $\kappa$-deformed bosons. All the quantities
$N_0^{^{\, {\scriptstyle \kappa}}}$, $U^{^{\,
{\scriptstyle\kappa}}}$, $T_c^{^{\, {\scriptstyle \kappa}}}$,
$C_V^{^{\, {\scriptstyle\kappa}}}$ of the 3-dimensional
$\kappa$-bosons here considered have finite value when
$0<\kappa<2/5$ (when $\kappa\geq 2/5$ the total kinetic energy
$U^{^{\, {\scriptstyle\kappa}}}$ diverges).

\newpage
\section{Figure Captions}
Fig.1. Temperature dependence of the particle ratio in the ground
state $N_0^{^{\, {\scriptstyle\kappa}}}/ N$ for $\kappa$-deformed
boson gas, with different $\kappa$ value. We remember that
$T_c^{^{\, {0}}}$ is the standard condensation temperature
obtained as $\kappa=0$. In the insert is plotted the $\kappa$
dependence of the $\kappa$-deformed and standard condensation
temperature ratio $T_c^{^{\, {\scriptstyle\kappa}}}/T_c^{^{\,
{0}}}$ as $\kappa$ takes values in the range $0<\kappa<2/5$.\\

Fig.2. Temperature dependence of specific heat $C_V^{^{\,
{\scriptstyle\kappa}}}/Nk_B$ of $\kappa$-deformed boson gas with
different $\kappa$ values in the range $0<\kappa<2/5$. $T_c^{^{\,
{0}}}$ is the standard condensation temperature obtained when
$\kappa=0$.
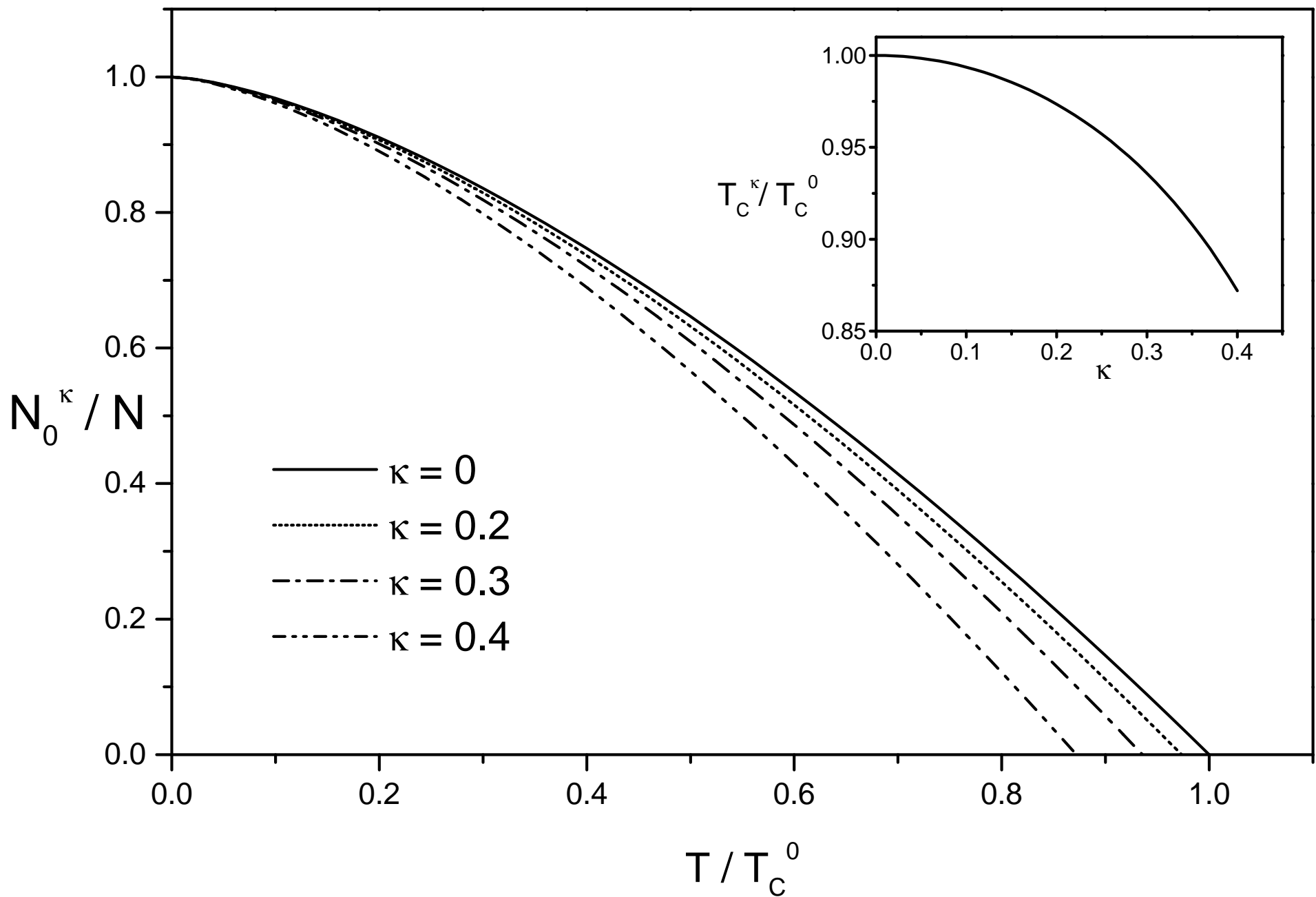
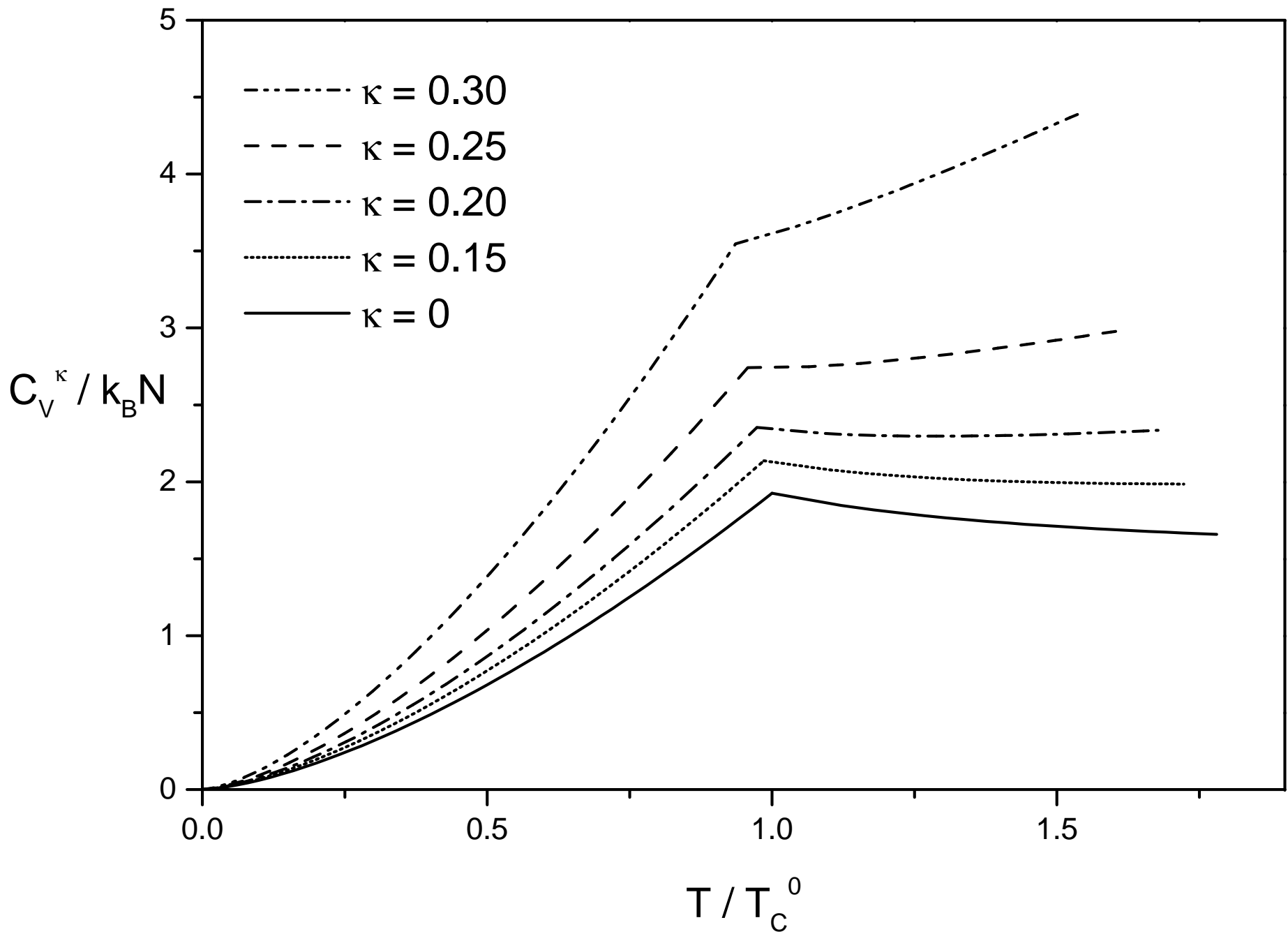
\end{document}